\documentstyle[prd,aps,floats]{revtex}  

\begin{document}  

\draft

\renewcommand{\topfraction}{0.99} 
\renewcommand{\bottomfraction}{0.99} 
  
  
\title{Inflation and Braneworlds: Degeneracies and Consistencies} 
\author{Greg Huey and James E.~Lidsey} 
\address{Astronomy Unit, School of Mathematical 
Sciences,   
Queen Mary, University of London, \\ 
Mile End Road, LONDON, E1 4NS, U.K.}

\date{\today} 
\maketitle 
\begin{abstract}  
Scalar and tensor perturbations arising 
in an inflationary braneworld scenario driven by a 
single scalar field are considered, where the bulk 
on either side of the brane 
corresponds to 
Anti--de Sitter spaces with different cosmological constants. 
A consistency relation between the two spectra is derived and 
found to have an identical form to that arising in standard 
single--field inflation based on conventional Einstein gravity. 
The dS/CFT correspondence may provide further insight into the 
origin of this degeneracy. 
Possible ways of lifting such a degeneracy are discussed. 
\end{abstract}   
  
  

\section{Introduction}

\setcounter{equation}{0}

The observational evidence for 
acoustic peak structure in the power spectrum of the 
cosmic microwave background (CMB) radiation 
continues to increase \cite{cmb}. The data is consistent 
with an initial spectrum of adiabatic scalar (density) perturbations 
generated in a spatially flat universe 
\cite{teg,efstathiou,hudod}. This 
provides strong support for the simplest class of inflationary 
models, where the accelerated expansion 
of the universe is 
driven by the potential energy arising from the 
self--interactions of a scalar `inflaton' field. (For recent 
reviews, see, e.g., Ref. \cite{ll}). 
Density perturbations in the post--inflationary 
universe are generated due to 
quantum fluctuations in this field as it rolls 
slowly 
down its potential \cite{perturbations}. 

In general, the spectrum of density perturbations 
is given by $A^2_S \propto k^{n_S-1}$, where 
the dependence of the amplitude, $A_S^2$, 
on comoving wavenumber, $k$, is parametrized 
in terms of the spectral index, 
$n_S =n_S(k)$. 
During inflation, a primordial spectrum of 
tensor (gravitational wave) perturbations is 
also generated quantum mechanically with 
amplitude $A^2_T \propto k^{n_T}$ \cite{starwave}. 
To lowest--order in the slow--roll approximation, 
the spectral indices, $\{ n_S ,n_T \}$,
have constant, but different, values \cite{llkcba}. 
However, the amplitudes of 
the two spectra on a given scale 
are not independent 
and are related through the so--called 
`consistency' relation\footnote{In this paper, 
we adopt the normalization conventions of Ref. \cite{llkcba} 
for the amplitudes of the perturbation spectra. 
For the tensor spectrum, this implies that 
$A_T^2 = \langle \varphi^2_0 \rangle /50$, 
where $ \langle \varphi_0^2 \rangle^{1/2} 
=2\kappa_4 (H/2\pi )$ is the amplitude of the 
quantum fluctuation in each of the polarization states, $\varphi_0$,
and $\kappa^2_4 = 8\pi/m^2_4$, where $m_4$ 
is the four--dimensional Planck mass. Units are 
chosen such that $\hbar = c =1$.}: 
\begin{equation}
\label{consistency}
\frac{A^2_T}{A^2_S} = -\frac{1}{2} n_T  .
\end{equation}

The primary 
importance of the consistency equation (\ref{consistency})
is that it reduces the number of independent inflationary 
parameters to three. 
More optimistically, since it does not depend  
on the specific functional form of the inflaton potential,
it may be regarded as 
a model--independent test of 
the single--field inflationary scenario, 
although the practical difficulties in measuring the relevant 
parameters are considerable. To date, 
the
contribution from tensor perturbations to the 
CMB power spectrum is constrained to 
be no more than 30 per cent \cite{teg}, but 
interest is growing in the possibility that forthcoming 
measurements of the CMB polarization 
may result in a direct detection 
of the gravitational waves through their contribution to the 
curl (B--mode) component of the polarization  
\cite{Bmode}. 
(For a recent survey of the observational status see Ref.  
\cite{bond}). 

From the theoretical perspective, 
there has been 
considerable interest in recent years in the possibility that our 
observable (four--dimensional) universe 
may be represented as a domain wall or membrane 
that is embedded in 
a higher--dimensional `bulk' spacetime
\cite{early,hw,lukas,RSII,kraus,ida,othermoves,quadratic}.
In particular, 
in the type II Randall--Sundrum model (RSII) \cite{RSII}, 
a domain wall is embedded in five--dimensional Anti--de 
Sitter (AdS) space. 
Corrections to Newtonian gravity are suppressed even though 
the fifth dimension is uncompactified
because the higher--dimensional 
geometry is non--factorizable \cite{RSII}. 

The 
motion of the brane through the
bulk space is interpreted as cosmic expansion or contraction
by an observer confined to the 
brane \cite{kraus,ida,othermoves}. The warping 
of the bulk on the brane 
leads to corrections to the standard Hubble law
and 
the Friedmann equation acquires a term that is quadratic in the 
energy density \cite{quadratic}. These corrections enhance the friction 
acting on the inflaton field at high energies, 
thereby extending the region of parameter space where inflation 
may proceed \cite{maartens,steep,gl,sahni}. 

Such effects also 
significantly modify the amplitudes and scale dependences of the 
scalar and tensor perturbations \cite{maartens,lmw}. Remarkably, 
however, it was recently shown that the 
corresponding consistency equation in the RSII scenario has 
precisely the form 
given in Eq. (\ref{consistency}) \cite{gl}. 
This results in a degeneracy between 
the predictions of the different scenarios. 
An important question that arises, therefore,
is whether 
this degeneracy between the consistency equations 
of the standard and braneworld inflationary scenarios 
is merely a coincidence, or whether it 
is indicative of a more universal relationship. 

In this paper we explore such a degeneracy further. 
One of the primary features of the RSII model is 
the ${\rm Z}_2$ reflection symmetry that is imposed 
on the fifth dimension of the bulk metric. We 
relax this assumption by considering
an asymmetric braneworld scenario, where the 
bulk space on either side of the brane is represented by different 
AdS spaces with different, negative cosmological constants. 
A natural mechanism for breaking the reflection symmetry 
in this way is to couple the brane to a four--form 
gauge field through a Wess--Zumino action \cite{carter}.
The cosmology of asymmetric braneworlds has been considered 
previously by a number of authors 
\cite{kraus,ida,carter,stw,ch,davis}. 
In this paper, we consider a scenario where the inflationary 
expansion of the 
braneworld is driven by a single, self--interacting scalar 
field that is confined to the brane. 
The effect of the ${\rm Z}_2$ symmetry 
on the form of the consistency equation can then
be determined. 

We begin in Section II by 
presenting the Friedmann equation for the scenario 
under consideration and proceed to derive the 
scalar perturbation amplitude
generated during an 
epoch of inflation. 
We derive the corresponding 
expression for the tensor perturbations
in Section III. In Section 
IV, it is shown that the consistency equation 
(\ref{consistency}) holds even when the 
reflection symmetry is not imposed. 
In Section V, the degeneracy is discussed 
within the context of the recently conjectured dS/CFT 
correspondence \cite{dscft}
and possible ways to lift the degeneracy 
are highlighted.

\section{Cosmological Field Equations and Scalar Perturbations}

\setcounter{equation}{0}

When deriving the Friedmann equation, it is convenient  
to choose coordinates in the vicinity of the 
brane such that the brane is stationary 
and the metric dependence on the fifth coordinate is separable, 
i.e., $ds^2_5=\gamma_{\mu\nu} (x^{\sigma}, 
y) dx^{\mu}dx^{\nu} + dy^2$, where 
the brane is located at\footnote{Upper case, Latin indices denote 
five--dimensional
variables, $A=(0, 1,\ldots, 4)$, lower case, 
Greek indices represent world--volume indices, $\mu =(0, \ldots, 3)$, 
and spatial indices on the domain wall 
are denoted by lower case, Latin indices, $i=(1,2,3)$.}  $y=0$.  
The effective Friedmann equation is then 
derived from the junction conditions \cite{israel}: 
\begin{equation}
\label{junction}
\left[ K_{\mu\nu} \right]  =-\kappa^2_5 \left( T_{\mu\nu}
-\frac{1}{3} T \gamma_{\mu\nu} \right)   ,
\end{equation}
where $[K_{\mu\nu}] = K_{\mu\nu} (0^+) -K_{\mu\nu} (0^-)$, 
$K^{AB} ={^{(4)}} \nabla^{A} n^{B}$ is the extrinsic curvature 
of the boundary with unit normal vector, $n^{A}$, $T_{\mu\nu}$
is the energy--momentum of the matter confined to the 
brane, $T=T^{\mu}_{\mu}$ is its trace, 
$\kappa^2_5 \equiv 8\pi G_5$ and $G_5$ is the five--dimensional 
Newton constant.

Since we focus on inflation of the braneworld, 
we consider the case where the 
bulk metric corresponds to pure AdS space
and the 
induced metric on the brane is  
the spatially flat, 
Friedmann--Robertson--Walker (FRW) metric.
It is assumed that the
brane energy--momentum tensor 
is given by the perfect fluid
form $T^A_B |_{\rm brane} = \delta (y) {\rm diag} 
(-\rho , p, p, p, 0)$, 
where $\rho$ and 
$p$ represent the energy density and pressure, 
respectively. We focus on the case where 
the energy--momentum is sourced by the tension 
of the brane, $\lambda$, and a single, 
self--interacting scalar field, $\phi$, with 
potential, $V(\phi )$, such that
$\rho = \rho_{\phi} +\lambda$
and $p=p_{\phi} -\lambda$, where
$\rho_{\phi} =\dot{\phi}^2/2 +V(\phi )$, 
$p_{\phi}=\dot{\phi}^2/2 -
V(\phi )$ and a dot denotes differentiation with 
respect to proper time on the brane. The 
spatial components of the junction conditions 
(\ref{junction}) then reduce to \cite{kraus,ida,stw,ch}
\begin{equation}
\label{junction1}
\left( \alpha_+ +H^2 \right)^{1/2} +\left( 
\alpha_- +H^2 \right)^{1/2} =
\frac{\kappa^2_5 \rho}{3}   ,
\end{equation}
where $\alpha_{\pm} \equiv -\kappa^2_5 \Lambda_{\pm}/6$,  
$\Lambda_{\pm}$ represent the bulk cosmological 
constants on either side of the brane,
$H=\dot{a}/a$ is the Hubble parameter
and $a(t)$ is the scale factor of the world--volume metric. 
Eq. (\ref{junction}) can be solved to yield  
the effective Friedmann equation on the brane \cite{kraus,ida,stw,ch}:
\begin{equation}
\label{friedmann}
H^2 =\frac{\kappa_5^4 \rho^2}{36} - \frac{1}{2} \left( \alpha_- +\alpha_+ 
\right) +\frac{9}{4 \kappa^4_5 \rho^2} \left( \alpha_- -\alpha_+ 
\right)^2  .
\end{equation}

Relaxing the ${\rm Z}_2$ symmetry 
results in the non--trivial 
third term on the right hand side of Eq. (\ref{junction1}).
It is interesting to note that when this term is present, 
Eq. (\ref{junction1}) is invariant under an 
infra--red/ultra--violet duality transformation on the 
total energy density, 
$\kappa^2_5 \rho \leftrightarrow 
9| \alpha_+ -\alpha_- |/(\kappa_5^2\rho)$. 
As in the symmetric scenario, 
the standard, linear dependence on the energy density 
may be recovered at low energy 
scales by
tuning the brane tension to cancel the effects of the negative cosmological 
constants. Dimensional reduction implies that the 
four-- and 
five--dimensional Newton constants are related by \cite{stw}
\begin{equation}
\label{fourfive}
\frac{\kappa^2_5}{\kappa^2_4} = \frac{1}{2} 
\left( \frac{1}{\sqrt{\alpha}_+} +\frac{1}{\sqrt{\alpha_-}} 
\right)  .
\end{equation}
In the symmetric limit, $\Lambda_{\pm} =\Lambda$, 
Eq. (\ref{friedmann}) reduces to \cite{quadratic}
\begin{equation}
\label{symfri}
H^2 =\frac{\kappa^2_4\rho_{\phi}}{3} \left[ 1+\frac{\rho_{\phi}}{2\lambda}
\right]  ,
\end{equation}
where the brane tension is tuned to satisfy 
$\kappa^2_5 \lambda = \sqrt{-6\Lambda }$. 

When Eq. (\ref{junction1}) is valid, 
the time--component of the junction 
condition (\ref{junction})
enforces covariant conservation 
of energy--momentum on the brane \cite{kraus}. 
This implies that
\begin{equation}
\label{scalareom}
\ddot{\phi} +3H\dot{\phi} +V' =0
\end{equation}
for a self--interacting scalar field, where a prime denotes $d/d\phi$, and
Eqs. (\ref{junction1}) and (\ref{scalareom}) then 
determine the classical dynamics of the asymmetric braneworld. 

We now proceed to determine the amplitude of the 
scalar perturbations that are 
generated during inflation of the asymmetric braneworld. 
We consider the era when the slow--roll approximation,
$|\dot{H} | /H^2 \ll 1$ 
and $|\ddot{\phi}| \ll H | \dot{\phi} |$, is valid. 
In general, the curvature perturbation on uniform density 
hypersurfaces is given by $\zeta = H\delta \phi /\dot{\phi}$ 
and this in turn 
is determined by the scalar field fluctuation, $\delta \phi$, 
on spatially flat hypersurfaces \cite{bst}. 
The perturbations generated from a single, self--interacting 
scalar field are adiabatic and 
conservation of energy--momentum then implies that $\zeta$ is conserved 
on large scales \cite{wands}. Consequently, the 
amplitude of a mode when it re--enters the Hubble radius after inflation 
is related to the curvature perturbation by 
$A^2_S = 4 \langle \zeta^2 \rangle /25$. 
The right--hand side of this expression is evaluated 
when the mode goes beyond the Hubble radius 
during inflation, i.e., when the comoving wavenumber 
is given by 
\begin{equation}
\label{kdef}
k(\phi) = a_e H(\phi ) \exp [ -N(\phi ) ]   ,
\end{equation}
where $a_e$ denotes the 
value of the scale factor at the end of inflation 
and $N =\ln a =\int dt H (t)$ 
represents 
the number of e--foldings between a scalar field 
value, $\phi$, and the end of inflation, $\phi_e$. 
Since the
field fluctuation at this epoch is
determined by the Gibbons--Hawking temperature of 
de Sitter space,
$\langle \delta \phi^2 \rangle = H^2/(4\pi^2)$, 
it follows that the scalar perturbation amplitude has the form 
\begin{equation}
\label{scalar}
A^2_S = \left. \frac{1}{25\pi^2} \frac{H^4}{\dot{\phi}^2}
\right|_{k=aH}
\end{equation}
Substitution of the scalar field equation (\ref{scalareom}) 
then relates the amplitude directly to the inflaton potential: 
\begin{equation}
\label{scalar1}
A_S^2 = \left. \frac{9}{25\pi^2} \frac{1}{V'^2} 
\left[  
\frac{\kappa_5^4 (V + \lambda )^2}{36} - \frac{1}{2} 
\left( \alpha_- +\alpha_+ 
\right) +\frac{9}{4 \kappa^4_5 (V + \lambda )^2} \left( \alpha_- -\alpha_+ 
\right)^2
\right]^3 \right|_{k=aH}
\end{equation}

\section{Tensor Perturbations}

\setcounter{equation}{0}

The calculation of the gravitational wave spectrum is more 
involved, because the tensor perturbations extend into the 
bulk.  
Langlois, Maartens and Wands \cite{lmw}
have considered the generation 
of gravitational waves in the symmetric RSII scenario 
in the limit where the world--volume of the brane 
corresponds to pure de Sitter space. This case arises 
when the scalar field is constant 
and represents a good approximation 
to a field that is rolling slowly down its potential. 
In this limit, 
the evolution equation for the tensor 
perturbations admits a separable solution and this allows an 
analytical expression for the amplitude 
on large scales to be derived. 

We briefly review the method of Ref. \cite{lmw}
and then extend the analysis to the asymmetric scenario. 
The unperturbed bulk metric in the ${\rm Z}_2$ symmetric 
model is written as
\cite{kaloper}
\begin{equation}
\label{bulkmetric}
ds^2_5 = {\cal{A}}^2(y) \left[ -dt^2 + a^2 (t) dx^2 \right] +dy^2  ,
\end{equation}
where ${\cal{A}} = (H/\alpha )
\sinh [\alpha (y_h -|y| )]$,  
the Cauchy horizons,  
$g_{00} (\pm y_h ) =0$, 
are located at $y=\pm y_h$, and the constant 
$\alpha =\kappa_4/\kappa_5 = (-\Lambda /6 )^{1/2}$ is determined by the 
bulk cosmological constant, $\Lambda$.
The perturbed metric is given by 
$ds^2_5 = {\cal{A}}^2 [-dt^2 + a^2 (\delta_{ij} +E_{ij})dx^i
dx^j] + dy^2$ and the 
metric perturbations, $E_{ij}$, are 
decomposed into Fourier modes
with amplitude,  $E(t,y; \vec{k} )$. 
The linearly perturbed junction conditions (\ref{junction}) 
then imply that 
\begin{equation}
\label{pjunction}
\left. \frac{dE}{dy} \right|_{y=0} =0
\end{equation}
in the absence of anisotropic 
stresses.
The equation of motion for the gravitational waves separates into 
`on--brane' and `off--brane' components
and the amplitude is expanded into eigenmodes, 
$E(t, y; \vec{k}) =\int dm \varphi_m (t; \vec{k}) {\cal{E}}_m (y)$,
where $m$ represents the separation constant. 
The general solution for the zero--mode is 
given by 
\begin{equation}
\label{zerogeneral}
{\cal{E}}_0 =C_1 + C_2 \int^y dy' \frac{1}{{\cal{A}}^4(y')}  ,
\end{equation}
where $C_{1,2}$ are constants. 
Although Eq. (\ref{zerogeneral}) diverges 
at the Cauchy horizons,
the boundary condition 
(\ref{pjunction}) removes the divergent part of the 
zero--mode, since it requires $C_2=0$. 
The 
general solution to the off--brane equation 
for the `light' modes ($m<3H/2$) remains divergent at the Cauchy horizon
even when Eq. (\ref{pjunction}) is satisfied. 
Consequently, 
these modes do not contribute to the spectrum of 
orthonormal modes that forms the basis of the Hilbert space
for the quantum field.
`Heavy' modes with $m>3H/2$ remain in the vacuum state during inflation. 

The solution 
to the on--brane equation for the zero--mode asymptotes to 
$\varphi_0 \rightarrow {\rm constant}$ at late times. Thus, the amplitude
of the zero--mode 
remains constant on super--Hubble radius scales as in the 
standard inflationary scenario.
The amplitude of the quantum fluctuations in this
mode is determined by deriving an 
effective, five--dimensional 
action for the tensor perturbations and integrating over the 
fifth dimension. Normalizing the action relative to the 
standard, four--dimensional result imposes 
the condition \cite{lmw}
\begin{equation}
\label{Znormalization}
2\int_0^{y_h} dy C_1^2 {\cal{A}}^2 
=1
\end{equation}
and this implies that $C_1=\sqrt{\alpha} F(x )$, where 
\begin{equation}
\label{Fdefine}
\frac{1}{F^2} = \sqrt{1+x^2} -x^2 {\rm sinh}^{-1} 
\left( \frac{1}{x} \right)
\end{equation}
and $x \equiv H/\alpha$. 
This normalization results in a 
four--dimensional action for the zero--mode 
that is formally equivalent to 
that of a massless scalar field in a FRW universe, but 
with an overall factor of $(8\kappa_5^2)^{-1}$ instead of the 
conventional factor of $(8\kappa^2_4)^{-1}$. 
Consequently, the standard four--dimensional 
results can be employed by viewing 
each polarization, $\varphi_0$,  
as a quantum field
evolving in a time--dependent potential. 
This implies that 
$\langle \varphi_0^2 \rangle^{1/2} = 2\kappa_5 (H/2\pi)$
and the tensor perturbation amplitude is 
therefore given by \cite{lmw}
\begin{equation}
\label{symmetrictensor}
A_T^2 
= \frac{C_1^2 \langle \varphi_0^2 \rangle}{50}   
=\frac{\kappa_4^2}{50\pi^2}H^2F^2  .
\end{equation} 

We now determine the gravitational wave amplitude 
when the ${\rm Z_2}$ symmetry is not imposed in the fifth 
dimension. The method of Ref. \cite{lmw}
may be employed directly, and we therefore 
omit many of the details.
The first question to address is the appropriate 
form of the bulk metric. 
In the asymmetric scenario, regions of two AdS spaces with 
different cosmological constants 
are effectively separated by the brane. Consequently, 
when the bulk metric is expressed in 
Gaussian normal coordinates, 
it is possible to consider cases where 
a Cauchy horizon exists on only one side of the 
brane. However, 
in the derivation of Eq. (\ref{junction1}), 
it was assumed implicitly that 
the unit normal vector, $n^A$, pointed towards a Cauchy horizon. 
(In the cases where this is not so, the signs of 
one or both of the terms on 
the left hand side of Eq. (\ref{junction1}) would change, 
thus requiring unphysical forms of matter). 

For consistency, therefore, we should consider 
the scenario where 
Cauchy horizons exist on both sides of the brane. 
The appropriate form for the unperturbed bulk metric 
is then given by Eq. (\ref{bulkmetric}), where
\begin{eqnarray}
\label{Apm}
{\cal{A}}(y)  = \left\{ \begin{array}{ll}
{\cal{A}}_+  (y) \equiv  {\rm cosech} (\sqrt{\alpha_+}y_{h_+}) 
\sinh ( \sqrt{\alpha_+} (y_{h_+} -y )), 
&  \mbox{if $y>0$} \\
{\cal{A}}_- (y) \equiv {\rm cosech} (-\sqrt{\alpha_-} y_{h_-}) 
\sinh( \sqrt{\alpha_-} (y-y_{h_-} )), 
& \mbox{if $y<0$ ,}
\end{array}
\right. 
\end{eqnarray}
where the Cauchy horizons are located at 
\begin{equation}
y_{h_{\pm}} = \pm \frac{1}{\sqrt{\alpha_{\pm}}} 
{\rm sinh}^{-1} \left( \frac{\sqrt{\alpha_{\pm}}}{H}
\right) 
\end{equation}
and $H$ is the four--dimensional Hubble parameter. 
The bulk metric is continuous 
at $y=0$, as required.
The locations of the Cauchy horizons are determined by the 
values of the bulk cosmological constants
and the two horizons 
are equidistant from the brane when the cosmological constants
have the same value. 
This implies that the ${\rm Z}_2$--symmetric scenario 
can be recovered in a smooth fashion in the limit that  
$\Lambda_+ \rightarrow \Lambda_-$. 
 
In general, the linearly perturbed 
junction conditions imply that 
\begin{equation}
\label{qjunction}
\left. \frac{dE}{dy} \right|_{y = 0^+} = 
\left. \frac{dE}{dy} \right|_{y = 0^-}
\end{equation}
in the absence of a reflection symmetry.
Thus, a given mode and its first 
derivative must be continuous across the brane, unless there exist
anisotropic stresses. (We assume such effects to be 
negligible in the present work).
The general solution for the zero--mode is now given 
in terms of the quadrature (\ref{zerogeneral}), where
Eq. (\ref{Apm}) is satisfied, and this diverges at 
the Cauchy horizon unless $C_2=0$, as in the symmetric model. 
Consequently, the $C_2$ part of the solution 
does not contribute to the quantum vacuum. 
This leaves the constant solution 
${\cal{E}}_0=J={\rm constant}$ as the physically 
relevant solution to the off--brane equation of
motion. Specifying this solution 
is equivalent to 
imposing 
the boundary condition $(dE/dy)|_{y=0^{\pm}} =0$ 
and results in a normalizable zero--mode. 
The asymptotic form of the 
light modes ${\cal{E}}_m$ may also be deduced 
and the non--trivial solutions, as well 
as the general solution, diverge at the Cauchy horizon. 

It follows from Eq. (\ref{Znormalization}) 
that the normalization of the zero--mode is sensitive to the 
location of the Cauchy horizon. 
In the asymmetric case, this condition
generalizes
to
\begin{equation}
\label{Jdefine1}\frac{1}{J^2} = \int^0_{y_{h_-}} dy {\cal{A}}^2_- (y) +
\int^{y_{h_+}}_0 dy {\cal{A}}_+^2 (y)
\end{equation}
and 
substituting the bulk metric (\ref{Apm}) and evaluating 
the integrals then implies that  
\begin{equation}
\label{Jdefine2}
\frac{1}{J^2} = \frac{1}{2\sqrt{\alpha_-} F^2(x_-)} 
+ \frac{1}{2\sqrt{\alpha_+}F^2(x_+)}  ,
\end{equation}
where $F=F(x_{\pm})$ is given in Eq. 
(\ref{Fdefine}) and 
$x_{\pm} \equiv H/\sqrt{\alpha_{\pm}}$.

Finally, integrating over the fifth dimension 
in the effective action for the tensor perturbations results 
in a four--dimensional action for the zero--mode, $\varphi_0$,
that is identical in form to that of the symmetric model. 
We conclude, therefore, 
that the gravitational wave spectrum generated in the asymmetric 
inflationary 
braneworld scenario is given by 
\begin{equation}
\label{tensor}
\left. A^2_T =\frac{\kappa^2_5}{50\pi^2} H^2 J^2
\right|_{k=aH}
\end{equation}
We refer to 
Eq. (\ref{Jdefine2}) as the correction function
for the gravitational waves. At low energies, $x_{\pm} \ll 1$ 
and
$J \rightarrow \kappa_4/\kappa_5$, implying that the standard expression 
is recovered in this limit.

\section{Consistency Equation}

\setcounter{equation}{0}

Eqs. (\ref{scalar1}) and (\ref{tensor}) 
represent the scalar and tensor perturbation spectra
for a given inflationary potential. 
Relaxing the ${\rm Z}_2$ reflection symmetry has 
resulted in 
both spectra 
receiving
significant modifications. 
Given the complicated functional 
form of the two spectra, it might be expected that 
the form of consistency equation (\ref{consistency}) 
would be altered by these additional terms. 

In the standard approach to deriving the consistency equation,  
we differentiate the tensor spectrum with 
respect to comoving wavenumber, $k$, and employ the condition 
$k=aH$ to relate a given scale to a particular value of 
the inflaton field \cite{llkcba}. To lowest--order in the 
slow--roll approximation, this implies that\footnote{In 
the following discussion, 
equality denotes equality at this level of the slow--roll approximation.} 
$d\ln k = Hdt \equiv dN$. The consistency equation then 
follows by substituting in the expressions for the amplitudes. 

On the other hand, it is more illuminating to 
first consider 
the nature of the consistency relation 
between the two spectra in the high energy limit, 
where $\kappa^4_5\rho^2 \gg 1$ 
$( x_{\pm} \gg 1)$. In this limit, 
the first term on the right hand side of Eq. 
(\ref{scalar1}) dominates. 
Thus, at high energies, both the 
Friedmann equation (\ref{friedmann}) and 
the scalar perturbation amplitude (\ref{scalar}) reduce to the 
corresponding equation arising in the 
symmetric model. Moreover, it follows from Eq. (\ref{Jdefine2}) 
that $J^2 \approx 3H/2$
in this limit, implying that the tensor amplitude 
also reduces to the symmetric limit. 
Thus, the degeneracy 
of the consistency equation is not lifted in the high--energy limit
and 
the tilt of the tensor spectrum is related to the ratio of 
the scalar and tensor amplitudes by Eq. (\ref{consistency}). 

Motivated by the above observation, we now 
consider the possibility that Eq. (\ref{consistency}) is 
valid for all energy scales. In this case, 
substituting Eqs. (\ref{scalar}) and (\ref{tensor}) 
into Eq. (\ref{consistency}) implies 
that 
\begin{equation}
\label{require1}
\frac{d \ln (HJ)}{d H} \frac{dH}{dN} = - \frac{\kappa^2_5}{18}
\frac{J^2V'^2}{H^4}
\end{equation}
and Eq. (\ref{require1}) can be simplified 
by expressing 
the scalar field equation (\ref{scalareom}) 
in the form
\begin{equation}
\label{require2}
\frac{dH}{dN} = - \frac{d H}{dV} \frac{V'^2}{3H^2}  .
\end{equation}
Substituting Eq. (\ref{require2}) into (\ref{require1}) 
then yields 
\begin{equation}
\label{require3}
H^4 \frac{dH}{dV} \frac{d (HJ)^{-2}}{dH}  = - \frac{\kappa_5^2}{3} .
\end{equation}
Eq. (\ref{require3}) 
represents a necessary condition that  
the effective Friedmann equation, $H=H(V)$, 
and 
correction function, $J=J(H)$, must satisfy for the 
consistency equation (\ref{consistency}) 
to remain valid. 

To proceed, it proves convenient to define the new 
variables
\begin{equation}
\label{defineu}
u_{\pm} \equiv {\rm sinh}^{-1} \left( 
\frac{\sqrt{\alpha_{\pm}}}{H} \right)  .
\end{equation}
It then follows from the definition (\ref{Fdefine}) 
that the correction function (\ref{Jdefine2}) 
may be expressed in the compact form
\begin{equation}
\label{compact} 
\frac{1}{H^2 J^2} = \frac{1}{4} \sum_{i=\pm }
\frac{1}{\alpha_i^{3/2}} \left[ \sinh \left( 2u_i 
\right) -2u_i \right] 
\end{equation}
and differentiating Eq. (\ref{compact}) 
with respect to the Hubble parameter implies that 
\begin{equation}
\label{d}
\frac{d(HJ)^{-2}}{dH} =-\frac{1}{H^2} 
\sum_{i=\pm }  \frac{1}{\alpha_i} \left( \sinh u_i 
\right) \left( {\rm tanh} u_i \right) .
\end{equation}
Eq. (\ref{require3}) may now be simplified by substituting 
in Eq. (\ref{d}): 
\begin{equation}
\label{further}
\frac{dH}{dV} \left[ {\rm sech} u_+ + {\rm sech} u_-  \right] 
=\frac{\kappa^2_5}{3}
\end{equation}
and the identity   
\begin{equation}
\frac{d}{dV} \left( \sqrt{\alpha_{\pm}} 
{\rm coth} u_{\pm} \right) = \frac{dH}{dV} {\rm sech} u_{\pm}
\end{equation}
then results in a further simplification:
\begin{equation}
\label{necessary}
\frac{d}{dV}  \left[ \sqrt{\alpha_-} {\rm coth} u_-
+\sqrt{\alpha_+} {\rm coth} u_+ \right] = \frac{\kappa^2_5}{3} .
\end{equation}

Thus, 
the consistency equation for the asymmetric braneworld 
inflationary scenario is given by 
Eq. (\ref{consistency}) {\em if} 
Eq. (\ref{necessary}) is satisfied. 
We may establish that this 
is indeed the case
by rewriting the junction condition 
(\ref{junction1}) in terms of the new variables 
(\ref{defineu}): 
\begin{equation}
\sqrt{\alpha_-} {\rm coth} u_- + \sqrt{\alpha_+} 
{\rm coth} u_+ =\frac{\kappa^2_5}{3} (V +\lambda)   ,
\end{equation}
where we have employed the slow--roll approximation. 
Since the tension of the brane, $\lambda$, is 
constant, we deduce immediately that Eq. (\ref{necessary}) 
follows as a direct consequence of the junction condition 
(\ref{junction1}), or equivalently, the effective 
Friedmann equation (\ref{friedmann}). 

\section{Discussion}

\setcounter{equation}{0}

The consistency equation (\ref{consistency}) 
has long been 
regarded as a strong observational signature 
of the standard, single--field inflationary scenario 
formulated within the 
context of conventional Einstein gravity \cite{llkcba}. 
We 
have considered the corrections that arise to the 
scalar and tensor perturbation spectra in a Randall--Sundrum
braneworld scenario, where the bulk cosmological constants on either 
side of the brane are different. 
We have found that the corresponding 
consistency equation 
for such an asymmetric braneworld model
is also given by 
Eq. (\ref{consistency}). 
This identifies a third class of inflationary models
where such a relationship between observable 
parameters arises and 
the result is surprising given that 
the gravitational physics and perturbation spectra 
are radically different in all three scenarios. 
In other models where the spectra differ from the standard expressions, 
such as in the warm inflationary scenario \cite{warm}
and models driven by higher--order terms in the 
curvature invariants \cite{higher}, 
the consistency equation is modified. 

In effect, the consistency equation is 
given by Eq. (\ref{consistency}) because 
the Hubble parameter, $H=H(V)$, and correction function, $J=J(H)$, 
satisfy the first--order differential equation
(\ref{require3}). The corrections to the scalar and tensor spectra 
in both the symmetric and asymmetric braneworld scenarios have 
precisely the necessary form for this equation to remain valid. 
It is not obvious {\em a priori} from 
Eqs. (\ref{scalar1}), (\ref{Jdefine2})
and (\ref{tensor}) that this should be the case. 

It is of interest to explore 
the degeneracy of the consistency equation 
from a theoretic viewpoint. 
Further insight may be gained by considering the
symmetric Randall--Sundrum model. By defining 
the new variables 
\begin{eqnarray}
\label{defineb}
b \equiv \frac{1}{2} \sinh^{-1} x \\
\label{definebeta}
\beta \equiv \kappa_4 \frac{d \phi}{dN}  ,
\end{eqnarray}
where $x$ is defined after Eq. (\ref{Fdefine}), 
it is possible to rewrite the 
cosmological field equations (\ref{symfri}) 
and (\ref{scalareom}) as a {\em first--order} 
system \cite{hl}
\begin{equation}
\label{dotb}
\dot{b} =- \left( \frac{3\kappa^2_4}{8 \lambda} \right)^{1/2} \dot{\phi}^2
\end{equation}
and 
\begin{equation}
\label{bprime}
\beta =-\left( \frac{8 \lambda}{3} \right)^{1/2} 
\frac{b'}{H}  .
\end{equation}

The 
correction function (\ref{Fdefine}) 
to the gravitational wave spectrum 
can then be expressed 
directly in terms of a quadrature with respect to the $b$--function: 
\begin{equation}
\label{Fquad}
\frac{1}{F^2} = -4 \sinh^2 2b\int \frac{db}{\sinh^3 2b} 
\end{equation}
and the scalar perturbations (\ref{scalar}) 
take the form
\begin{equation}
\label{betascalar}
A_S^2 =\frac{\kappa^2_4}{25\pi^2} \frac{H^2}{\beta^2}   .
\end{equation}
By employing the definitions (\ref{kdef}), 
(\ref{defineb}) and (\ref{definebeta}) 
and substituting Eqs.  
(\ref{symmetrictensor}), (\ref{bprime}) and (\ref{betascalar}) into 
Eq. (\ref{Fquad}), we find that 
\begin{equation}
\frac{1}{A^2_T} = 2 \int \frac{d \ln k}{A^2_S}
\end{equation}
when the slow--roll approximation, $|\dot{H}|/H^2 \ll 1$ 
and $| \ddot{\phi} | \ll H |\dot{\phi} |$, 
is obeyed. 
Thus, 
differentiation with respect to comoving 
wavenumber recovers the consistency equation (\ref{consistency}).

The degeneracy between the
consistency 
equations of the standard and symmetric braneworld scenarios 
arises because the observable quantities, 
$\{ (A_T^2/A_S^2) , n_T \}$, 
in the latter case 
are independent of the brane tension. 
(The standard expressions for the perturbation spectra are 
formally recovered in the limit $\lambda \rightarrow \infty$).
The variables 
$\{ b , \beta \}$ 
play a central role in establishing this independence. 
Moreover, introducing these variables 
allows the second--order field 
equations (\ref{symfri}) and (\ref{scalareom}) 
to be expressed as a coupled, 
first--order system (\ref{dotb}) and (\ref{bprime}), where 
$b$ is a monotonically {\em decreasing} 
function of cosmic time when the null energy condition is 
satisfied \cite{hl}. This is analogous 
to the flow equations recently 
considered within the context of the conjectured dS/CFT correspondence
\cite{dscft,lsl,argurio}
and motivates us to explore a physical interpretation of these parameters 
within such a context. 

The dS/CFT correspondence states that 
pure quantum gravity in de Sitter (dS)
space 
has a dual description in terms of 
a conformal field theory (CFT), where the latter is 
located on the Euclidean boundary of de Sitter space at future 
infinity \cite{dscft}. Although still at the conjectural level, 
this correspondence establishes a `dictionary' relating 
variables in the bulk to those of the boundary theory. 

In a field theory with an operator, ${\cal{O}}$, and coupling parameter, 
$g$, the coupling is constant at the classical level, but renormalization
introduces a scale--dependence. The breaking of 
scale invariance is parametrized 
by the $\beta$--function,  
$\beta \equiv \partial g/\partial \ln \mu$, where 
$\mu$ is the renormalization group (RG) scale. 
In the 
dS/CFT correspondence, a scalar field is dual to the operator, 
${\cal{O}}$.  
The symmetries of de Sitter space, and 
consequently the scale invariance of the CFT, 
are broken when the field rolls slowly down its potential \cite{lsl}. 
By analogy with the AdS/CFT correspondence
\cite{adscft}, the value of the scalar 
field is identified with the field theory coupling and 
the RG scale with the scale factor of the universe, i.e., 
$g =\kappa_4 \phi$ and $\mu \propto a$
\cite{lsl}. 
Thus, we may identify Eq. (\ref{definebeta}) as the 
$\beta$--function of the quantum theory expressed in terms of the bulk 
variables. This first--order equation then
describes the RG flow between the fixed points, $\beta =0$. 

Another quantity of key importance is the central charge
($c$--function). 
This parametrizes the number of degrees 
of freedom of the CFT and decreases along the RG flow to the 
infra--red 
fixed point. (In the dS/CFT correspondence, this limit corresponds to the 
past). 
A natural candidate for the $c$--function 
has been proposed in standard inflationary cosmology, 
$c \equiv (\kappa^2_4 H^2)^{-1}$ \cite{dscft}. 
Adopting such an expression
in the case of the braneworld implies that
the $c$--function is related to the $b$--function 
(\ref{defineb}) such that
$c =(3/4 \pi \lambda \kappa^2_4) {\rm cosech}^2 (2b)$. 
This decreases monotonically as we go back in time 
by virtue of Eq. (\ref{dotb}). 
Thus, the functions (\ref{defineb}) and (\ref{definebeta}) 
admit a physical interpretation in terms of the $\beta$-- and 
$c$--functions of the dual three--dimensional Euclidean field 
theory and the relative
amplitudes of the perturbation spectra can be expressed 
directly in terms of 
these variables: 
\begin{equation}
\label{directly}
\frac{A_T^2}{A_S^2} = \frac{1}{2} \beta^2 F^2 [c]  .
\end{equation} 
Since the dS/CFT correspondence introduces a new 
perspective on inflationary cosmology, it would be interesting to extend this 
preliminary analysis further along the lines outlined in Ref. 
\cite{lsl} for standard inflationary cosmology. 
 
The degeneracy of the 
consistency equation (\ref{consistency}) 
has a number of significant implications. 
In particular, it implies that observations 
of the CMB may not be
able to discriminate between conventional and braneworld inflation 
\cite{lt}
and this leads us to consider how 
the degeneracy might be lifted. 

It should be emphasized that the degeneracy arises in 
the predictions of the {\em primordial} perturbations,
but  
the bulk space
may influence the evolution of these perturbations. 
In particular, 
the derivation of Eq. (\ref{consistency}) neglected 
the backreaction 
of the metric fluctuations in the fifth dimension. This 
is a consistent approximation 
when the scalar perturbations are perturbations 
of a homogeneous distribution of matter on the brane, as is the 
case considered above 
\cite{maartens}. More 
generally, however, the backreaction induces a 
non--trivial Weyl curvature in the bulk and  
this manifests itself as a non--local source of 
energy--momentum in four dimensions 
\cite{gordonmaartens}.  
The background dynamics is altered by such a source and it would 
be interesting to incorporate these effects into the analysis. 
Furthermore, 
we have focused on a 
specific realization of the braneworld scenario, where the bulk space is 
conformally flat and 
the degeneracy may be broken  by relaxing 
this condition. This could be achieved, for example,
by introducing a scalar field into the bulk. 

The results discussed above hold only to lowest--order 
in the slow--roll approximation. To next--to--leading order, the 
consistency equation of the standard scenario receives 
corrections \cite{ckll}: 
\begin{equation}
\label{hc}
n_T = -2 \frac{A^2_T}{A^2_S} \left[ 1- \frac{A^2_T}{A^2_S} 
+ (1-n_S) \right]   .
\end{equation}
Significantly, these corrections
do not depend on the tilt of the tensor spectrum
and it is therefore important to 
establish whether similar terms arise in the braneworld 
scenario. Eq. (\ref{hc}) was derived by 
performing a general expansion about the 
exact perturbation spectra that are generated during power law inflation. 
This particular 
model is exactly solvable because the kinetic and potential 
energies of the scalar field redshift at the same rate. 
The corresponding model in the braneworld scenario was recently 
found \cite{hl} and, in principle, could serve as a basis 
for addressing higher--order effects in this class of models. 

It is known that
the consistency equation (\ref{consistency}) 
relaxes to an inequality, $A^2_T/A_S^2 \le -n_T/2$, 
in multiple--field inflationary models \cite{mul}.
Isocurvature (entropy)  
perturbations can also be generated when more than one scalar field 
is present. Recently, it was shown that for a general 
two--field model,  
the cross correlation between the adiabatic and isocurvature  
modes modifies the consistency 
equation such that $A^2_T/A^2_S=-n_T(1-r_C^2)/2$, where 
$r_C$ is determined by the cross--correlation power spectrum \cite{bmr}. 
It is possible, therefore,  that the degeneracy between the single field 
consistency equations might be lifted by considering
isocurvature perturbations in a two--field braneworld scenario. This would
require a detailed analysis and derivation of the evolution 
equations for the perturbations. 

In conclusion,  
there exists a surprising degeneracy between 
the observational predictions of different inflationary 
scenarios based on conventional Einstein  gravity and 
the braneworld picture. 
This provides us with 
potentially crucial 
observational insight into the gravitational physics of 
extra dimensions.

\acknowledgments

GH is supported by the Particle Physics and 
Astronomy Research Council. JEL is supported by 
the Royal Society. We thank D. Wands for helpful discussions.

\end{document}